\documentclass{epl}
\newcommand{\bbox}[1]{{\bm #1}}

\title{Caustics in turbulent aerosols}
\author{M. Wilkinson\inst{1} \and B. Mehlig\inst{2}}
\institute{
  \inst{1} Faculty of Mathematics and Computing,
  The Open University, Walton Hall, Milton Keynes, MK7 6AA, England\\
  \inst{2} Theoretical Physics, Physics and Engineering Physics, Gothenburg
  University/Chalmers, Gothenburg, Sweden
}


\begin{document}

\maketitle

\begin{abstract}
Networks of caustics can occur in the distribution of particles
suspended in a randomly moving gas. These can facilitate coagulation
of particles by bringing them into close proximity, even in cases
where the trajectories do not coalesce. The long-time
morphology of these caustic patterns depends upon the Lyapunov
exponents $\lambda_1$, $\lambda_2$ of the suspended particles,
as well as the rate $J$ at which particles encounter caustics.
We develop a theory determining the quantities $J$, $\lambda_1$,
$\lambda_2$ from the statistical properties
of the gas flow, in the limit of short correlation times.
\end{abstract}

Aerosols are usually unstable systems, in that the suspended 
particles eventually coagulate. Understanding the process giving rise
to
this coagulation, and determining  the time scale over which it occurs
are important questions in describing any aerosol system.
If the gas phase does not have macroscopic motion, the coagulation
may be effected by diffusion of the suspended particles, or
(if the suspended particles are of a volatile substance) 
by Ostwald ripening.
The coagulation process can be greatly accelerated if the aerosol
undergoes macroscopic internal motion. Ultrasound,
for example,
has been used to accelerate coagulation in aerosols \cite{Goo59}.
Turbulent flow could also play a role in the coagulation 
of suspended particles; this could be relevant in the 
coalescence of visible moisture into rain droplets 
\cite{Sha99}.

If suspended particles are simply advected in an incompressible
flow, their density remains constant. Inertial effects
are therefore required for coagulation, unless the flow exhibits significant 
compressibility. In earlier work \cite{Wil03,Meh03} we discussed
the motion of inertial particles in a random velocity
field. We showed that there is a phase where the trajectories 
of the particles coalesce, so that arbitrarily small particles 
coagulate. In the limit where the correlation time $\tau$ of the 
flow approaches zero, this path-coalescing phase only
exists when the velocity field is predominantly 
potential flow (such as the flow due to sound waves) \cite{Meh03}. 

Turbulent fluid flow is, however, expected to be predominantly solenoidal, 
and it is of interest to find alternative mechanisms
of coagulation which operate outside the path-coalescence phase.
\begin{figure}[bt]
\centerline{\includegraphics[width=12cm,clip]{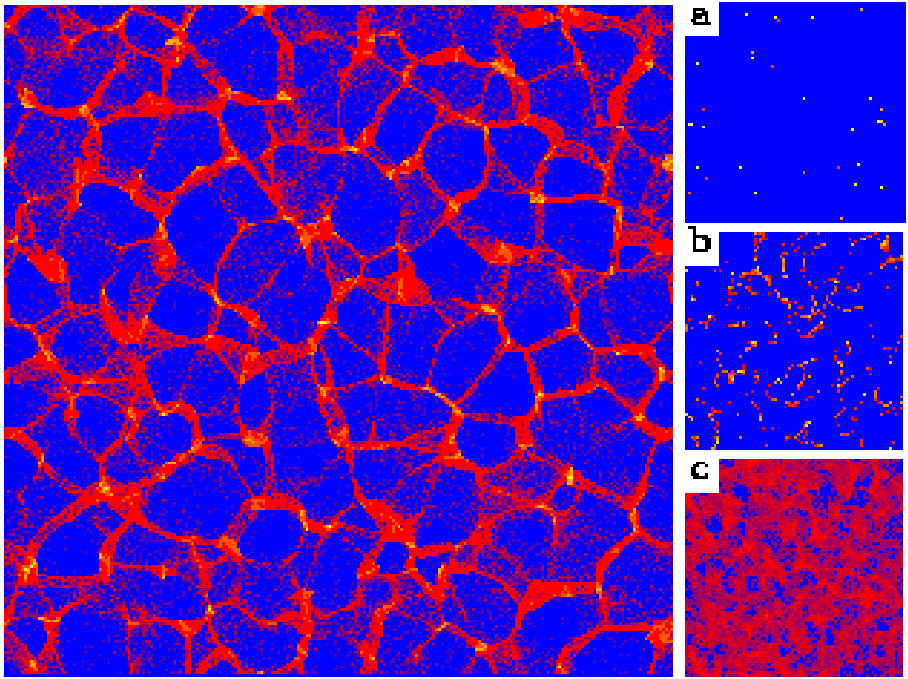}}
\caption{\label{fig: 1} Distribution of inertial particles 
suspended in a randomly moving fluid 
(blue corresponds to lowest density,
yellow to highest). The initial distribution is a random scatter.
The large panel shows caustics at short time. Panels (a)-(c)
show the long-time behaviour. In all cases, the region is the
unit square, the mean particle density is $2.5\times 10^{5}$, $m=1$, 
and there is potential flow (with parameters 
$\xi=0.03$, $\sigma=0.01$, $\delta t=0.05$, see text).
Main panel: $\gamma=0.53$, $t=5$, (a): $\gamma=1.18$, $t=500$, 
(b): $\gamma=0.72$, $t=125$,
(c): $\gamma=0.21$, $t=125$.
The three cases correspond to: (a) $\lambda_2 < \lambda_1 < 0$, (b) $\lambda_1 > 0$,
$\lambda_1+\lambda_2 < 0$, and (c) $\lambda_1 >0$, $\lambda_1+\lambda_2
>0$, see text.}
\end{figure}
Here we describe an
alternative mechanism facilitating coagulation, 
illustrated in Fig.~\ref{fig: 1}:
we show the distribution of particles suspended in a randomly
moving gas (the equations of motion and statistics of the flow
field are given by eqns.~(\ref{eq: 1}) to (\ref{eq: 3}) below). 
The large panel shows the distribution of particles after a 
short time, starting from a random scatter with uniform density. 
The particles cluster onto a network of caustic lines, 
analogous to the networks of optical caustics that can be 
seen on the bottom of a swimming pool \cite{Ber81}.
This phenomenon is a new mechanism by 
which aerosol particles are brought into close proximity.
The remaining parts of 
Fig.~\ref{fig: 1} show the distribution of 
particles after a long time, in three different cases: part (a)
shows the path-coalescence phase where the trajectories
condense onto points. Parts (b) and (c) show two cases
where there is no path coalescence, but a steady state 
with significant inhomogeneities of density due to caustics:
these have very different morphologies, depending on the parameter
values, as we shall show.

Fig.~\ref{fig: 1} is surprising because
it is be expected
that random movement of uniformly distributed particles would
leave the distribution uniform. 
The following questions 
naturally arise. First, why do the particle trajectories 
coalesce into points in Fig.~\ref{fig: 1}(a)? 
This phenomenon was first noted
in \cite{Deu85} and subsequently analysed in detail
in \cite{Wil03,Meh03}
(c.f. also the theory developed at the end of this paper).
Secondly, why and how does the caustic pattern develop? We will
explain why the caustics form, and relate the pattern to 
optical caustic networks in swimming pools. Third, why do 
the morphologies of the structures seen in the steady state
at long times differ? We will argue that the different 
morphologies of the caustic patterns are related to
three parameters: the rate $J$ at which any given
particle crosses a caustic and 
the Lyapunov exponents $\lambda_1$ and $\lambda_2$ of the flow 
(with $\lambda_1>\lambda_2$). 
We present a new theory
for these parameters in the 
limit where the correlation time of the random flow is short.
Finally, it is natural
to ask how the density fluctuations are quantified.
We will show that the probability distribution function
for the particle density has two regimes of algebraic decay.
The formation of caustics causes particles to pass through regions
of greatly increased density, where coagulation by contact interaction
is much more likely to occur. In this paper we confine ourselves 
to describing the caustics, and do not model the coagulation process.

There is a large literature on light-intensity fluctuations
in twinkling starlight, and other cases of propagation
through random media. Such problems are different because the 
finite wavelength of the light causes
the fluctuation distribution due to catastrophes 
to be cutoff at high intensities \cite{Ber81}.
Caustics also occur in Hamiltonian dynamics, and their
observation in electron flows is a topic of current
research \cite{Top03}, where the statistics
of caustic formation are of interest \cite{Kap02}. 
Our system differs from the Hamiltonian case:
the drag force on the particles is assumed to
be given by Stokes's law.  
Because of the Stokes damping, the density correlation function 
reaches a statistically stationary state, as opposed to
the Hamiltonian case.
Although our results are described in the context of aerosols, the same
principles apply to suspensions of particles in liquids and to tracer
particles used to investigate turbulent flows.
Because our objective is to explain the theoretical 
principles as clearly as possible, we confine discussion to 
one or two spatial dimensions.

We consider small spherical non-interacting particles of mass $m$, radius
$a$, in a random flow with velocity field $\bbox{u}(\bbox{r},t)$
and viscosity $\eta$. 
We assume that the drag force
on the particles is given by Stokes's law. Neglecting
displaced-mass effects, the equation of motion is
\begin{equation}
\label{eq: 1}
\ddot{{\bbox{r}}}=-\gamma (\dot{\bbox{r}}-{\bbox{u}})
\end{equation}
where $\gamma =6\pi \eta a/m$ and $\bbox{r}$ is the particle position.
This model is widely used in studies of particles suspended 
in turbulent fluids (see \cite{Sig02}, \cite{Bec03} for descriptions
of recent numerical investigations of this model), and it is surprising
that the occurrence of caustics was not noted earlier.
The random driving
force on the particles, $\bbox{f}=m\gamma\bbox{u}$, is conveniently
described by two scalar potentials $\phi$ and $\psi$:
\begin{equation}
\label{eq: 2}
\bbox{f}(\bbox{r},t)=\nabla \phi(\bbox{r},t)
+\nabla \wedge \hat{\bm n}_3\psi(\bbox{r},t)
\end{equation}
(where $\hat{\bm n}_3$ is a unit vector perpendicular to the plane):
the scalar fields $\phi$ and $\psi$ generate, 
respectively, the potential and solenoidal components of the flow.
Here we assume that $\phi$ and $\psi$ are independent, with
$\langle\phi\rangle=\langle\psi\rangle=0$ and 
$\langle \psi^2\rangle=\alpha^2\langle \phi^2\rangle$ for some constant 
$\alpha$ (angular
brackets denote expectation values). Also, we assume that 
the correlation function of $\psi$ has the same form
as that of $\phi$:
\begin{equation}
\label{eq: 3}
\langle \phi(\bbox{r}+\bbox{R},t_0+t)\phi(\bbox{r},t_0)\rangle=C(R,t)\,,
\end{equation}
where $R=\vert \bbox{R}\vert$, and $C(R,t)$ has correlation length 
$\xi$ and correlation time $\tau$. 
The theory is readily extended to more general statistics.

Our two-dimensional numerical simulations are performed
in the limit of small $\tau$, using a model discretised
in time with a small time step $\delta t \gg \tau$: the impulse
\begin{equation}
\label{eq: 12}
{\bm f}_n({\bm r}) =\int_{n\delta t}^{(n\!+\!1)\delta t}\!\!
{\rm d}t'\ {\bm f}({\bm r}_{t'},t')
\end{equation}
 at time $n\,\delta t$ is taken to
be of the form (\ref{eq: 2}) in terms of scalar fields $\phi_n(\bbox{r})$
and $\psi_n(\bbox{r})$ satisfying
$\langle \phi_n(\bbox{r})\phi_{n'}(\bbox{r}')\rangle
=\sigma^2\, \xi^2\, \delta
t\,\exp(-|\bbox{r}-\bbox{r}'|^2/2\xi^2)\delta_{nn'}$
and $\langle \psi_n(\bbox{r})\psi_{n'}(\bbox{r}')\rangle 
=\alpha^2\langle \phi_n(\bbox{r})\phi_{n'}(\bbox{r}')\rangle$.

\begin{figure}[b]
\centerline{\includegraphics[clip]{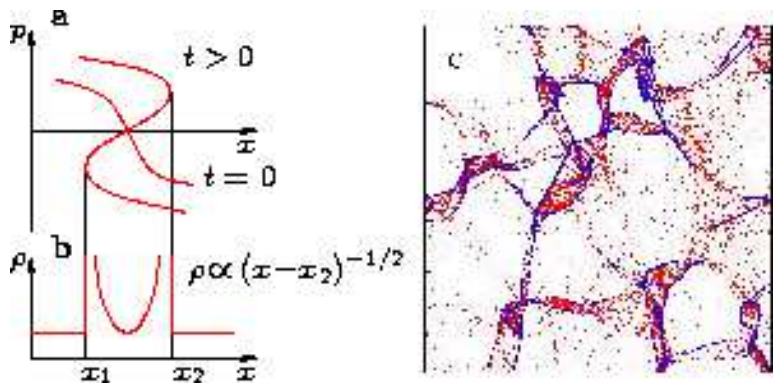}}
\caption{\label{fig: 2} 
(a) Particles are distributed on a phase-space manifold,
shown here as a phase curve in a one-dimensional section. 
The phase curve develops folds. (b) 
The particle density diverges on caustics, the projections
of the folds. The caustics are created in pairs, with a high density
of particles between each pair. (c) The particle distribution is 
shown in red, and the corresponding caustic curves are plotted as blue
lines. The parameters are the same as for the largest panel of 
Fig.~\ref{fig: 1}.}
\end{figure}
The origin of the caustics is most easily understood
by a one-dimensional example. Fig.~\ref{fig: 2} (a)
is a schematic plot of the momentum $p$ of a particle at
position $x$. At time $t=0$, this is a single-valued 
function, but at a later time this function may develop
a pair of folds because the faster particles overtake
the slower ones. If the density
of particles is smoothly distributed along the line in phase-space, 
the projected density in coordinate space, shown in part (b), 
is singular at a pair of caustics, which are the projections of each fold.
We see that the caustics are created in pairs, and that initially
there is a high density of particles between the caustic lines,
as well as a divergent density on the caustic itself.
Fig.~\ref{fig: 2} (c) shows a segment of the actual
caustic pattern. The caustic lines cannot abruptly end, but
caustics can join at cusps \cite{Ber81}, some of which are visible in the
figure. Some regions of high density do not have an associated
pair of caustic lines. This is because the phase-space manifold 
is nearly perpendicular to the coordinate-space plane, but it has
not yet folded over. At short times our caustic networks are equivalent 
to the lines of bright light observed on the bottom
of a swimming pool on a sunny day when the water surface is disturbed.
The morphology of the swimming-pool caustics is discussed by Berry
\cite{Ber81}.

There are important differences between optical realisations of caustic
patterns behind randomly refracting screens and the caustics in 
particle density. The former become more confused as the 
distance from the screen is increased, whereas our patterns reach
a statistically steady state at large times. Although more caustics are
created by folding, the damping term (with coefficient $\gamma$
in eq.~(\ref{eq: 1}) ) implies
that momentum differences decrease, so that the caustics become progressively
more tightly folded.

The examples in Fig.~\ref{fig: 1}(b) and (c) show that the morphology of 
the patterns in this steady state depends upon the statistics of the 
random velocity field. It is desirable to understand the differences
between these figures.

We consider the behaviour of three particles, a reference
particle and two nearby ones 
separated  by $\delta \bbox {r}$ and $\delta\bbox{r}'$. 
The evolution of these small
increments is described by the Lyapunov exponents, $\lambda_1$ and
$\lambda_2$, characterising the
time-dependence of the length $X(t)=\vert\delta\bbox{r}(t)\vert$
of one of the separation vectors, and 
of the area
of a parallelogram spanned by two vectors, 
$A(t)=\vert \delta \bbox{r}(t)\wedge \delta \bbox{r}'(t)\vert$.
These are defined by
\begin{equation}
\label{eq: 4}
\lambda_1\!=\!\lim_{t\to \infty}\!{1\over{t}}\!
\log_{\rm e}\bigg\vert{X(t)\over{X(0)}}\bigg\vert\,,\,
\lambda_1\!+\!\lambda_2\!=\!\lim_{t\to \infty}\!{1\over{t}}\!
\log_{\rm e}\bigg\vert{A(t)\over{A(0)}}\bigg\vert \ .
\end{equation}
\begin{figure}
\centerline{\includegraphics[width=10.5cm,clip]{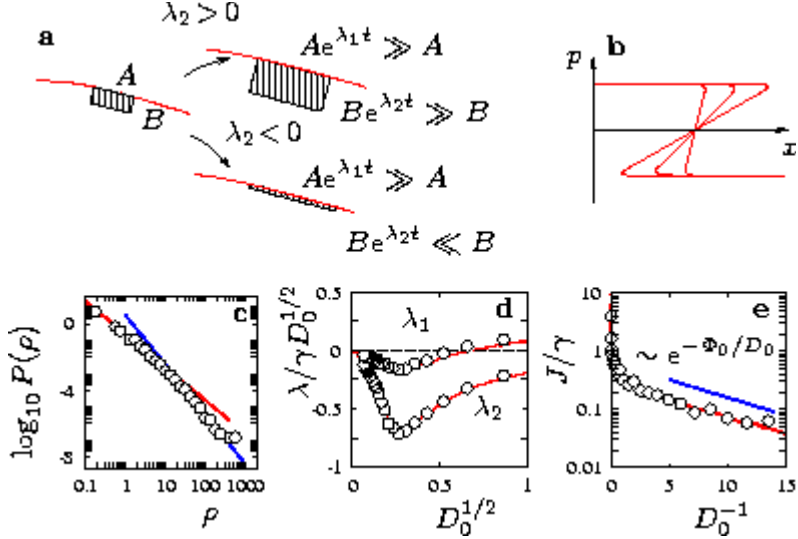}}
\vspace*{2mm}
\caption{\label{fig: 3} (a):
The Lyapunov exponents $\lambda_1$ and $\lambda_2$ determine the
particle-density fluctuations.
(b): Schematic illustration of the model for the folding process leading
to eq.~(\ref{eq: 5}).
(c): Density fluctuations in a one-dimensional
model ($\xi = 0.05$, $m=1$, $\tau=1$, $\sigma=10^{-3}$, $\gamma=0.04$,
averaged over 25 times ranging from $40\tau$ to $10^3\tau$): 
$P(\rho)$ shows
algebraic decay,
with a region of $\rho^{-2}$ decay followed by a $\rho^{-3}$
tail.
(d): Lyapunov exponents $\lambda_1$ and $\lambda_2$
for $\Gamma=1/3$; simulations of Langevin 
equations (lines) and integration of equation of motion ($\circ$). 
(e): Corresponding rate of crossing caustics.}
\end{figure}
Now we use the Lyapunov exponents to gain an understanding
of the patterns in Fig.~\ref{fig: 1}.
When $\lambda_1<0$, almost all infinitesimal line segments 
contract to a point, and the trajectories of particles coalesce
as shown in Fig.~\ref{fig: 1}(a). This was the principle 
used to explain the coalescence transition, discussed 
in \cite{Wil03} and \cite{Meh03}. Next consider what happens
if $\lambda_1$ is positive, so that the trajectories do not
coalesce into points. In this case we consider a small
element of area on a caustic line, 
extended in the direction of the caustic [Fig.~\ref{fig: 3}(a)].
This element is expected to be stretched along the direction of
the caustic, because we assume $\lambda_1>0$. If $\lambda_2$ is also 
positive, the line width of the element will also expand in
the direction perpendicular to the caustic, and the density
fluctuation resulting from the caustic becomes weaker as time
proceeds: this is illustrated by Fig.~\ref{fig: 1}(c).
If $\lambda_2<0$, the line
element contracts in the direction perpendicular to the caustic, 
and if in addition $\lambda_1+\lambda_2<0$, 
the concentration of particles on the caustic
increases. This case is illustrated by Fig.~\ref{fig: 1}(b), which
shows very narrow caustic lines.
The increase in particle density on the caustic lines cannot proceed 
without limit: the caustic line stretches and folds, until the particles
aligned along the caustic have become indistinguishable from points
randomly scattered in the plane. In this case, the caustic also disappears,
although this happens much more slowly than when $\lambda_1+\lambda_2>0$.

A useful quantitative understanding of the steady state
can be gained by considering the probability distribution of the 
particle density, $P(\rho)$. The particle density in the vicinity
of a caustic (at $x_0$, aligned with the $y$ axis, say) 
has a divergence of the form $\rho(x,y)\sim (x-x_0)^{-1/2}$.
This indicates that $P(\rho)\sim \rho^{-3}$ for sufficiently
large $\rho$. However, numerical evaluations of $P(\rho)$ 
indicates that $P(\rho)\sim \rho^{-2}$ over a significant range
of values, only conforming to the $\rho^{-3}$ prediction
for very large $\rho$. 
This surprising observation is a consequence of the 
nature of the folding process which generates the caustics,
and we give a brief heuristic argument which illustrates the principle.
Consider the schematic diagram in Fig.~\ref{fig: 3}(b), which
shows a simplified model for the evolution of a fold which forms
at $t=0$, with a depth which is proportional to time $t$.
The corresponding density distribution at time $t$ is 
$P(\rho,t)=tf(\rho t)$ for some function $f$. The pattern 
contains folds at different stages of their evolution, and the 
observed probability distribution function $P(\rho)$ is assumed
to be a time-average of $P(\rho,t)$, between $t=0$ and
some cutoff $t_0$ (which can be interpreted as the time
between the formation of folds). According to this model,
\begin{equation}
\label{eq: 5}
P(\rho)={1\over {t_0}}\int_0^{t_0}\!\!\!{\rm d}t\ P(\rho,t)
={1\over {\rho^2t_0}}\int_0^{\rho t_0}\!\!\!{\rm d}x\ x f(x)\sim \rho^{-2}
\end{equation}
(we assume that the integral of $xf(x)$ converges to a constant
as its upper limit approaches infinity). 
The model is not applicable at very short times,
where the fold resembles the generic type illustrated in Fig.~2(a).
Fig.~\ref{fig: 3}(c) illustrates the crossover from $\rho^{-2}$
to $\rho^{-3}$ decay in a numerical evaluation of $P(\rho)$.

We have argued that the morphology of the steady state
caustic patterns can be understood qualitatively if 
we calculate three parameters, namely the rate $J$ at 
which particles cross caustic lines, and the Lyapunov
exponents $\lambda_1$ and $\lambda_2$. We will now show
how these three parameters may be determined quantitatively.
The most complete results are available in the limit where the
correlation time $\tau$ is very short.

We linearise eq. (1), and obtain 
$\delta \dot{\bbox{r}}=\delta \bbox{p}/m$, $\delta \dot{\bbox{p}}
=-\gamma\delta \bbox{p}+{\bf F}(t)\delta \bbox{r}$, where ${\bf F}$
is a matrix with elements 
$F_{ij}(t)=\partial f_i(\bbox{r}(t),t)/\partial r_j$. 
Consider now the motion of three particles,
one reference particle and two nearby ones,
separated by $(\delta \bbox{r},\delta \bbox{p})$ and 
$(\delta \bbox{r}',\delta \bbox{p}')$ from the reference trajectory. 
The angle $\delta \varphi$ separating the vectors $\delta \bbox{r}$
and $\delta \bbox{r}'$ is very small, and the lengths of these vectors, 
$X$ and $X'$ respectively, are initially equal. We also assume
that the vectors $\delta \bbox{p}$ and $\delta \bbox{p}'$ are 
initially separated by a small angle, of order $\delta \varphi$.
We now make a change of coordinates 
from 
$\delta \bbox{r}$, $\delta \bbox{p}$, $\delta \bbox{r}'$, 
$\delta \bbox{p}'$ to the set $X,X',\theta,\delta \varphi,Y_1,Y_2,Z_1,Z_2$
\begin{eqnarray}
\label{eq: 6}
\delta \bbox{r}=X\hat{\bbox{n}}_\theta
&\ ,\ &
\delta \bbox{r}'=X'\hat{\bbox{n}}_{\theta+\delta \varphi}
\nonumber \\
\delta \bbox{p}=X(Y_1\hat{\bbox{n}}_\theta+Y_2\hat{\bbox{n}}_{\theta+\pi/2})
&\ ,\ &
\delta \bbox{p}'=
X'[(Y_1\!+\!Z_1\delta \varphi)\hat{\bbox{n}}_{\theta+\delta \varphi}\!+\!(Y_2+Z_2\delta \varphi)
\hat{\bbox{n}}_{\theta+\frac{\pi}{2}+\delta \varphi}]
\end{eqnarray}
($\hat{\bbox{n}}_\theta$ denotes a unit vector in 
two dimensions with direction $\theta$).
We expect that $X$ may increase or decrease, $\delta \varphi$
decreases with probability one (because random linear mapping of any
two vectors 
results in vectors becoming aligned), $X'/X$ remains close to unity, 
$\theta $ will become uniformly
distributed on $[0,2\pi]$, and that $Y_1$, $Y_2$, $Z_1$ and $Z_2$
approach a stationary distribution. The length of a vector separating
two nearby points is $X$, 
so that $\lambda_1=\langle \dot X/X\rangle$.
The area spanned by the two vectors is $\delta A\sim XX'\delta \varphi\sim
X^2\delta \varphi$, so that $\lambda_2+\lambda_1=
\langle \delta \dot \varphi/\varphi\rangle+2\langle \dot X/X\rangle$
(here we used the fact that $\delta \varphi \ll 1$). 

We find the equations of motion
for these variables. For $X$ and $\delta \varphi$ we find
$\dot X=Y_1X/m$ and $\delta \dot \varphi=Z_2\delta \varphi/m$, so that
the Lyapunov exponents are
\begin{equation} 
\label{eq: 7}
\lambda_1=\langle Y_1\rangle/m\,,\,\,\lambda_2=\lambda_1+\langle Z_2\rangle/m\,.
\end{equation}
For $\theta$, we find $\dot \theta=Y_2/m$, and conclude that $\theta$
executes a random walk, becoming uniform on $[0,2\pi]$ as expected.
For the remaining variables we find the following equations, 
where $\hat{\bbox{n}}_1=\hat{\bbox{n}}_\theta$, 
$\hat{\bbox{n}}_2=\hat{\bbox{n}}_{\theta+\pi/2}$,
$F'_{ij}(t)=\hat{\bbox{n}}_i.{\bf F}(t)\hat{\bbox{n}}_j$
\begin{eqnarray}
\label{eq: 8}
\dot Y_1 &=&-\gamma Y_1 +(Y_2^2-Y_1^2)/m
+ F'_{11}\\
\dot Y_2 &=& -\gamma Y_2 -2Y_1 Y_2/m 
+ F'_{21}
\nonumber \\
\dot Z_1 &=&-\gamma Z_1 -2\big( Z_1 Z_2/2 +Y_1 Z_1 -Y_2 Z_2\big)/m
+F'_{21} + F'_{12}
\nonumber \\
\dot Z_2 &=&-\gamma Z_2 -2\big(Z_2^2/2 +Y_1 Z_2 +Y_2 Z_1\big)/m
-F'_{11} + F'_{22}
\nonumber
\ .
\end{eqnarray}
In the limit where the correlation time of the random velocity
field is short, we can approximate eqns.~(\ref{eq: 8}) by a 
set of coupled Langevin equations. We write these in a dimensionless 
form by introducing dimensionless variables $x_i$, such that 
$(Y_1,Y_2,Z_1,Z_2)=m\gamma (x_1,x_2,x_3,x_4)\equiv m\gamma \bbox{x}$, 
$t'=\gamma t$ and find
\begin{equation}
\label{eq: 9}
{\rm d}{\bm x} = {\bm v}({\bm x}) {\rm d}t' + {\rm d}{\bm w}
\end{equation} 
where $v_1 =-x_1+(x_2^2-x_1^2)$, 
$v_2 =  -x_2-2 x_1 x_2$,
$v_3=-x_3-x_3 x_4 -2(x_1 x_3 -x_2 x_4)$,
$v_4 = -x_4-x_4^2-2(x_1 x_4 + x_2 x_3)$,
$\langle {\rm d}w_i\rangle=0$, 
$\langle {\rm d}w_i{\rm d}w_j\rangle = 2 D_{ij} {\rm d}t'$.  
The diffusion matrix with elements $D_{ij}$ is
\begin{equation}
\label{eq: 10}
{\bf D} = D_0
\left(
\begin{array}{cccc}
1 & 0                   &0     & \scriptstyle -(\Gamma+1)/2\\[0mm]
0 & \scriptstyle \Gamma & \scriptstyle (\Gamma+1)/2     & 0\\[0mm]
0 & \scriptstyle (\Gamma+1)/2  & \scriptstyle \Gamma+1  &0 \\[0mm]
\scriptstyle -(\Gamma+1)/2 & 0 &   0& \scriptstyle \Gamma+1
\end{array}
\right)
\end{equation}
where
\begin{equation}
\label{eq: 11}
D_0\!=\!{1\over{2m^2\gamma^3}}\int_{-\infty}^\infty\!\!\!\! {\rm d}t\ 
\langle F_{11}(t)F_{11}(0)\rangle
\ ,\ 
\Gamma={1+3\alpha^2\over{3+\alpha^2}}\,.
\end{equation}
In our simulations,
$D_0 = (3+\alpha^2)\sigma^2/(2m^2\gamma^3\xi^2)$.
The three parameters describing caustic formation
are obtained from the stationary state of the Langevin process
(\ref{eq: 9}): the rate $J$ at which a representative particle
passes through a caustic is the same as the frequency with
which the area of the parallelogram spanning the two vectors 
$\bbox{r}$ and $\bbox{r}'$ becomes equal to zero, or equivalently 
the rate at which $\delta \varphi$ passes through zero. An equivalent 
condition is that the trajectory in the $(Z_1,Z_2)$ plane goes to infinity
(reappearing from the reflected direction):
$J=\gamma j$, where $j$ is the 
rate at which $x_4$ escapes to infinity.
Fig.~\ref{fig: 3}(d) shows a comparison between a direct
evaluation of the Lyapunov exponents (using a method
described in \cite{Eck85}), and a simulation using
eq.~(\ref{eq: 9}).

Eq.~(\ref{eq: 9}) is equivalent to a Fokker-Planck equation
for the distribution $P({\bm x},t)$, namely
${\partial P}/{\partial t'} = 
\nabla.[-{\bbox{v}}P+{\bf D}\nabla P]$.
Considering a steady-state solution in the limit $D_0 \to 0$, we
find that a WKB ansatz, of the form 
$P(\bbox{x})=\exp[-\Phi(\bbox{x})/D_0]$, is appropriate.
We therefore expect that the escape current vanishes exponentially
as $D_0 \to 0$, being of the form $j\sim j_0\exp(-\Phi_0/D_0)$
(where $j_0$ may have an algebraic dependence on $D_0$). 
Fig.~\ref{fig: 3}(e) shows that the caustic formation rate does
vanish exponentially as $D_0\to 0$, with action $\Phi_0\approx 0.14$.

%\acknowledgements


\begin{thebibliography}{0}
\bibitem{Goo59} G. L. Gooberman, {\sl Ultrasonics: Theory and Application}, 
Hart Publications: New York, (1969).
%
\bibitem{Sha99} R. A. Shaw, {\sl Annu. Rev. Fluid Mech.}, {\bf 35}, 183,
(2003).
%
\bibitem{Wil03} M. Wilkinson and B. Mehlig, {\sl Phys. Rev. E}, {\bf 68}, 
040101, (2003).
%
\bibitem{Meh03} B. Mehlig and M. Wilkinson, {\sl Phys. Rev. Lett.}, 
{\bf 92}, 250602, (2004).
%
\bibitem{Ber81} M. V. Berry, {\sl Singularities in Waves}, Les Houches 
Lecture Series Session XXXV, eds. R. Balian, M. Kl\' eman and J-P. Poirier, 
North Holland: Amsterdam, 453-543, (1981).
%
\bibitem{Deu85} J. Deutsch, {\sl J. Phys.}, {\bf A18}, 1457, (1985).
%
\bibitem{Top03} M. A. Topinka, R. M. Westervelt, and E. J.
Heller, Physics Today, Dec. 2003, p. 47.
%
\bibitem{Kap02} L. Kaplan, {\sl Phys. Rev. Lett.}, {\bf 89}, 184103, (2002).
%
\bibitem{Sig02} H. Sigurgeirsson and A. M. Stuart, {\sl Phys. Fluids}, 
{\bf 14}, 4352-61, (2002).
%
\bibitem{Bec03} J. Bec, {\sl Phys. Fluids}, {\bf 15}, L81-84, (2003). 
%
\bibitem{Eck85} J. P. Eckmann and D. Ruelle, Rev. Mod. Phys. {\bf 57}, 617-656,
(1985).
%
\end{thebibliography}
\end{document}